\documentstyle[12pt,aasms4]{article}

\def\t0{\theta_{\circ}}

\def\be{\begin{equation}}
\def\en{\end{equation}}

\def\msun{M_{\sun}}

\def\lsun{L_{\sun}}

\begin{document}

\title {A dust disk surrounding the young A star HR4796A}
\author{Ray Jayawardhana\altaffilmark{1,3},
Scott Fisher\altaffilmark{2,3},
Lee Hartmann\altaffilmark{1},
Charles Telesco\altaffilmark{2,3},
Robert Pina\altaffilmark{2},
and Giovanni Fazio\altaffilmark{1}}
\altaffiltext{1}{Harvard-Smithsonian Center for Astrophysics, 60 Garden St., Cambridge, MA 02138; Electronic mail: rjayawardhana@cfa.harvard.edu}
\altaffiltext{2}{Department of Astronomy, University of Florida, Gainesville, FL 32611;
Electronic mail: telesco@astro.ufl.edu}
\altaffiltext{3} {Visiting Astronomer, Cerro Tololo Interamerican Observatory,
National Optical Astronomy Observatories, which is operated by the Association
of Universities for Research in Astronomy, Inc. (AURA) under cooperative
agreement with the National Science Foundation.}

\begin{abstract}
We report the codiscovery of the spatially-resolved dust disk of the Vega-like
star HR 4796A.
Images of the thermal dust emission at $\lambda = 18 \, \mu$m show an elongated
structure approximately 200 AU in diameter surrounding the central A0V star.
The position angle of the disk, $30^{\circ} \pm 10^{\circ}$, is consistent
to the position angle of the M companion star, $225^{\circ}$, suggesting
that the disk-binary system is being seen nearly along its orbital plane.
The surface brightness distribution of the disk is consistent with the
presence of an inner disk hole of approximately 50 AU radius,
as was originally suggested by Jura et al. on the basis
of the infrared spectrum. HR 4796 is a unique system among the 
Vega-like or $\beta$ Pictoris stars
in that the M star companion (a weak-emission T Tauri star)
shows that the system is relatively young, $\sim 8 \pm 3$~Myr.
The inner disk hole may provide evidence for coagulation of
dust into larger bodies on a timescale similar to that suggested for planet
formation in the solar system.  
\end{abstract}


\section{Introduction}

Planets are thought to form from the dusty disks that are the
remnants of star formation (Shu, Adams, \& Lizano 1987; Lissauer \& Stewart 1993).
The timescale for planet formation is uncertain, but is currently
thought to be roughly 10 Myr, as judged from both astronomical
and solar system constraints (Strom et al. 1989; Strom, Edwards,
\& Skrutskie 1993; Podosek \& Cassen 1994).
This suggests that the young T Tauri stars of ages of 
1 Myr, which frequently have optically-thick, actively-accreting disks
(Bertout 1989; Hartmann \& Kenyon 1996),
represent a stage prior to the main epoch of planet formation.
In contrast, the dust envelopes or disks
of the so-called Vega-like objects (Backman \& Paresce 1993),
main sequence stars
whose ages could be as large as 1 Gyr, are thought to be much
more evolved; most of the dust has coagulated into planets
or planetesimals, and the remaining dust in the optically-thin disk is
continually replenished by collisions between larger bodies
(Nakano 1988; Backman \& Paresce 1993).

To date the dust cloud of one Vega-like object, the nearby main-sequence A star
$\beta$ Pictoris, has been imaged in scattered light with sufficient resolution
and sensitivity to show disk structure (Smith \& Terrile 1984);
the dusty disk
is also observable in spatially-resolved mid-infrared thermal dust emission
(Telesco et al. 1988; Backman, Gillett, \& Witteborn 1992; Lagage \& Pantin 1994;
Pantin, Lagage, \& Artymowicz 1997).

Here we report imaging observations of thermal dust emission
for another Vega-like A star, HR 4796A.
This star has a dust excess spectrum qualitatively similar to
that of $\beta$~Pic, and its dust luminosity to stellar luminosity ratio
is about twice that of $\beta$~Pic ($L_{IR}/L_{*} = 5 \times 10^{-3}$;
Jura 1991; Jura et al. 1993; Gillett 1986).  
Our $18 \, \mu$m images show a disk, apparently seen nearly edge-on and with
a position angle nearly aligned with the M star companion.
The disk has an observed outer radius of 110 AU,
comparable to that of T Tauri stars (Dutrey et al. 1995).  
This result has been found independently and simultaneously 
by Koerner et al. (1998).

\section{Observations and Results}

HR 4796 was observed in March 1998 with the 4-m Blanco telescope   
at Cerro Tololo Interamerican Observatory using the OSCIR mid-infrared
camera. OSCIR uses a 128$\times$128 Si:As Blocked Impurity Band (BIB)
detector developed by Boeing. On the CTIO 4-m telescope, OSCIR has
a plate scale of 0.183''/pixel, which gives a field of view of 
23''$\times$23''. Our observations were made using the standard chop/nod
technique with a chopper throw of 25'' in declination. Images of HR 4796
were obtained in the K(2.2 $\mu$m), N(10.8 $\mu$m), and IHW18(18.2 $\mu$m)
bands, and flux calibrated using the standard star $\gamma$ Cru. Total
on-source integration times for HR 4796 were 648 seconds in K, 1800 
seconds in N, and 1800 seconds in IHW18. Additional information on 
OSCIR is available on the World-Wide Web at www.astro.ufl.edu/iag/.

In the IHW18 band, the dust disk surrounding HR 4796A
is clearly resolved along the major axis and marginally resolved
along the minor axis (Figure 1). The disk appears nearly edge-on in our images;
considering our point-spread function, we estimate an
angular diameter of 3'', consistent with the 5'' upper
limit from Jura et al. (1993). We measure a total flux of $1.1 \pm 0.15$ Jy
in a 3'' aperture around HR 4796A. The position angle of the
disk is $30^{\circ} \pm 10^{\circ}$. 

In the N-band, HR 4796A is marginally resolved along the direction of 
the disk's major axis, and not resolved perpendicular to it (Figure 2).
This result is as expected given our sensitivity limits,
and given that the dust excess at that wavelength is small in 
comparison to the stellar photospheric emission (Jura et al. 1998).  
Our N-band flux measurement of $244 \pm 25$ mJy in a 3'' aperture
agrees well with that of Jura et al. (1998).
We place a 3$\sigma$ upper limit of 23 mJy at N on the flux from B,
consistent with the 65 mJy limit reported by Jura et al. (1998).

In the K-band, both HR 4796A and B are point sources with FWHM not 
appreciably larger than that of the standard star.

\section{Discussion}

The infrared excess of HR 4796A can be fit by a blackbody
at 110 K; the absence of shorter-wavelength emission apparently
requires a depletion or absence of dust inside of $\sim 30$ AU 
(Jura et al. 1993).  To see whether our images are consistent
with the inference of an inner disk hole,
we have constructed preliminary dust disk models
for HR 4796A. These models assume that the disk is optically and geometrically thin.
We also adopt power-law distributions of surface density and temperature. 
Using a power-law dependence of the dust opacity on frequency at 
wavelengths $\geq 10 \mu$m, we require that the models reproduce the
infrared spectrum (Jura et al. 1998).  The $18.2 \mu$m stellar flux is
extrapolated from the $10.8 \mu$m flux. Finally, 
the {\it Hipparchos} distance of 67.1$^{+3.5}_{-3.4}$ pc is used,
consistent with the main sequence spectral type of A0V
($L_* \sim 21 \lsun$, effective temperature $\sim$~9500K; Jura et al. 1998).

Because we only marginally resolve the disk along the minor axis, we have chosen
to compare the models only with the strip surface brightness distribution, summed
along the minor axis.  We formally adopt an inclination angle of $\cos i = 0.3$ to
be consistent with the observations, but since the disk is optically thin, this
parameter makes no difference to the strip surface brightness.

Within the context of this simple model, an inner hole or region of dust depletion 
is needed to avoid having too large a central peak in the emission.
As shown in Figure 3, models with constant surface density require an inner hole 
radius $40~{\rm AU} \lesssim R_i \lesssim 80$~AU.
The surface density distribution as a function of cylindrical radius $R$
is not well constrained; models with
$\Sigma \propto R^{-p}$, with $p \sim 0 - 2$, are consistent with the data (Figure 3).

Although our model results are not unique, we note that the derived parameters
are very similar to those inferred for the $\beta$~Pic disk by Pantin et al. (1997).
By modelling their 10$\mu$m data, Pantin et al. find a relatively constant surface
density between about 50 and 100 AU, with rapid decreases to shorter and larger
radial distances.  Thus, the size of inner hole inferred here for HR 4796A is similar to
the inner hole radius estimated for $\beta$~Pic.

All of the models require dust temperatures of $\approx 80 - 90$~K at 100 AU,
which is much higher than the blackbody grain temperature of 60 K at this distance
(Jura et al. 1988).  This suggests that the grains must have reduced efficiency
of emission at mid-infrared wavelengths.  We have therefore adopted a power law
opacity of $\kappa_{\nu} \propto \nu^1$, so that the models all have a 
radiative equilbrium temperature distribution $T \propto R^{-0.4}$.
  
Calculations suggest that circumstellar disks will be truncated
by the tidal effects of a companion star in circular orbit 
at approximately 0.9 of the average Roche lobe radius
(Artymowicz \& Lubow 1994),
which for the estimated mass ratio of this object should
be slightly more than half the orbital radius
(Papaloizou \& Pringle 1977).  In the present case
this suggests the disk should be truncated at about 270 AU
rather than the observed 110 AU.  Our simple model (Figure 3) suggests 
that $18 \mu$m image does not trace emission from the outermost, coldest regions of the 
disk very well (the outer radius of our models is 200 AU), so that
the disk may extend well beyond what we can detect at 18~$\mu$m.
Alternatively, it may be that the orbit of the companion 
is eccentric, and that the M star is currently near apastron.

At an angular separation of approximately 7.7 arcsec, the companion
(spectral type M2.5; Stauffer et al. 1995 $=$ SHB)
star must be a physical partner because the stars have not changed
their relative positions for more than 60 yr (Jura et al. 1993).
As discussed by SHB, the difficulty in assigning an age to the M star derives
mostly from calibrations of theoretical evolutionary tracks. In effect,
SHB use the distance above the Pleiades main sequence
in the HR diagram to provide the calibration, and find an age of $8 \pm 2$~Myr.
Here we consider the effect of the new {\it Hipparcos} distance on
this result.  We adopt the temperature scale and bolometric corrections
adopted by Brice\~no et al. (1998), along with the I (Cousins) magnitude
reported by Jura et al. (1993) to determine an effective temperature
of $3620 \pm 60$~K and luminosity $0.11 \pm 0.02 \lsun$ for HR 4796B.
Then, using the $Hipparchos$ distance, and the D'Antona \& Mazzitelli (1994)
evolutionary tracks (for which the above temperature scale is valid),
the resulting age is $8 \pm 3$~Myr and the estimated mass is 
$0.38 \pm 0.05 \msun$.  This is virtually identical to the SHB result;
the (small) change in distance is compensated for by the change in
bolometric correction.  As noted by SHB, the measurement of the strong 
Li absorption line at 670.7 nm constrains the age to be less than $\sim 9-11$~Myr 
for this mass range (see, e.g., D'Antona \& Mazzitelli 1994).  
A lower limit to the age of a few Myr is indicated by
the isolated location of HR 4796, since most
stars of comparable or smaller ages are found in regions of
molecular clouds and substantial interstellar dust extinction (Leisawitz,
Bash, \& Thaddeus 1989).

The inner disk hole has been attributed variously to either the tidal
effects of  a ``sweeper'' planet or brown dwarf, or to inward migration
of dust due to Poynting-Robertson (PR) drag, followed by sublimation
of ice grains near 30 AU (see Jura et al. 1998).
Our observations do not constrain these possibilities.  The observations
suggest a flatter surface density distribution outside of the hole
than in $\beta$ Pic, which may indicate qualitatively different conditions.
We note that our upper limit of 23 mJy to the flux at 10.8 $\mu$m from the M star
is consistent with the predicted photospheric flux of $\sim 15$~mJy,
implying little if any dust emission from a potential disk
around the companion.  This is consistent with the M star being a weak-emission
T Tauri star, which are not thought to be accreting from inner disks (Bertout 1989). 

In the case of most Vega-like systems, it is assumed that the dust grains
responsible for the far-infrared excess emission must be continually
replenished by collisions and sublimation of
larger bodies, because the timescales for orbital decay by 
Poynting-Robertson (PR) effect
and sublimation of small bodies near the inner disk edge are smaller than the 
stellar main sequence lifetimes (Nakano 1988; Backman \& Paresce 1993).
The HR 4796 system is so young that this conclusion may not be applicable.
Jura et al. (1998) suggested that the grains need to be larger than
$100 \, \mu$m in order not to be removed from the disk by PR drag.
However, radiative equilibrium for such large grains should be close to the blackbody 
case, resulting in lower dust temperatures and making it much more
difficult to explain the size of the 18~$\mu$m image, as noted above.
It is possible
that small grains might be produced by collisions of larger bodies, as
suggested for $\beta$~Pic (see Backman \& Paresce 1993); alternatively,
if sufficient gas remains in the disk, as might be expected for a young
system, PR drag can be overcome by gas drag.  Attempts
to measure gas directly in this disk should be made to see if the amounts
are as small as found for $\beta$ Pic (e.g., Ferlet, Hobbs, \& Vidal-Madjar
1987).

Finally, it may be that there is not a universal evolutionary timescale
for protoplanetary disks, especially when the influence of companion stars
is taken into account.  Highly-simplified viscous disk models which are
consistent with T Tauri star mass accretion rates suggest that the decay
in disk mass as a function of time might be relatively slow for isolated stars,
leaving 0.001 - 0.01 $\msun$ of gas in a disk of 1000 AU radius at an age of 10 Myr
(Hartmann et al. 1998).  In contrast,
the mass of circumstellar dust in HR 4796A is estimated to be only $6 \times 10^{26}$ -
$6 \times 10^{27}$~g (Jura et al. 1995, 1998).
However, the presence of a companion star at 500 AU is likely to dramatically
accelerate the depletion of the disk due to accretion onto the central star
(Papaloizou \& Lin 1995).  The disk 
around HR 4796 can serve as a valuable laboratory for understanding disk 
evolution and planet formation.

We wish to thank the staff of CTIO for their outstanding support.  The research
at the University of Florida was supported by NASA, NSF, and the University
of Florida.  The research at CfA was supported by NASA grant NAG5-4282 and the
Smithsonian Institution.

\newpage

\centerline{\bf Figure Captions}

\bigskip
\bigskip

Figure 1 (color plate) - False-color image of HR4796 disk in the
IHW18 (18.2$\mu$m) band with surface brightness contours overlaid. 
The contours are at 50, 75, 100, 125, 150, and 175 mJy/arcsec$^2$.
The positions of star A and star B are marked with crosses, as determined
from reference star offsets and the K-band image.

Figure 2 - Surface brightness contour plots of the N-band, {\it upper left}, 
and IHW18, {\it upper right}, images of HR 4796A smoothed with a five 
pixel Gaussian. In the N image, the lowest contour is at 6.6 mJy/sq.arcsec 
and the contour interval is 9 mJy/arcsec$^2$. In the IHW18 image,
the lowest contour is at 44 mJy/arcsec$^2$ and the interval is
27 mJy/arcsec$^2$.  In lower panels we show the corresponding
point spread functions, with contouring at the same fractional values
of the peak emission as in upper panels. The lowest contour levels 
have been chosen to avoid the low-level, extended emission from the 
third diffraction ring in the PSF.

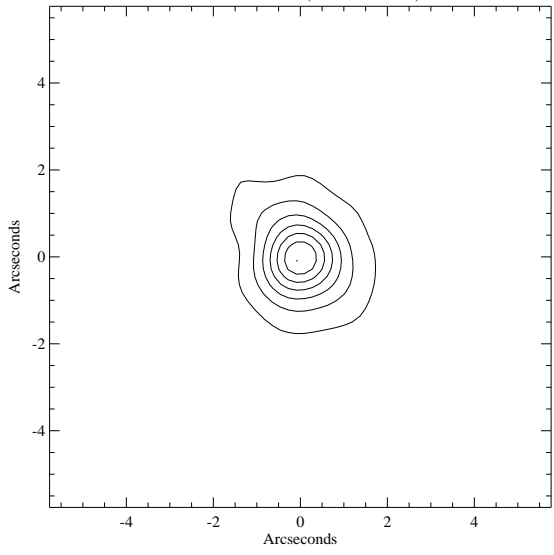
Figure 3 - The observed intensity of HR 4796A at 18.2 $\mu$m
summed along the minor axis (solid line), compared with 
simple disk models (dashed lines).
The observations have been smoothed by three pixels (0.54 arcsec).
The disk model is geometrically and optically thin; the dust opacity is
proportional to $\kappa_{\nu} \propto \nu^1$; the temperature distributions
are $T \propto R^{-0.4}$, and scaled to match the {\it IRAS} fluxes
(Jura et al. 1998); and the surface density has a power law form, $\Sigma \propto R^{-p}$.
The models have the indicated inner disk radii and values of $p$.  
All models have been convolved with a Gaussian point spread function with 
the same FWHM as the observed PSF.

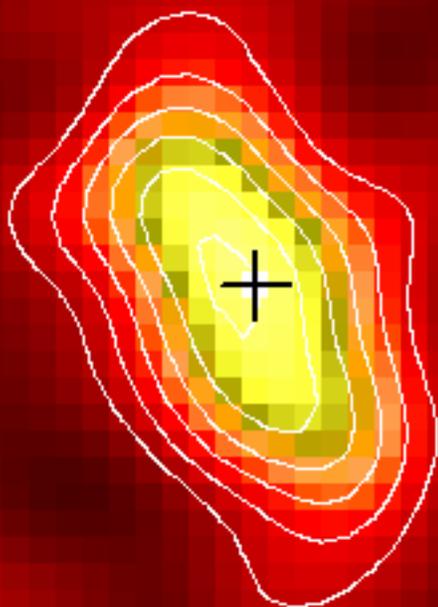
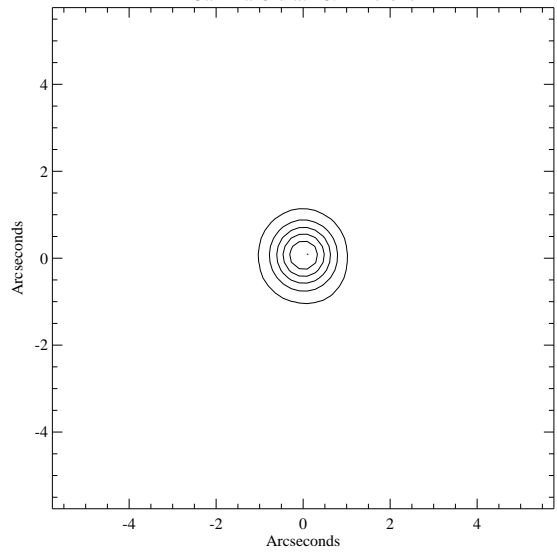
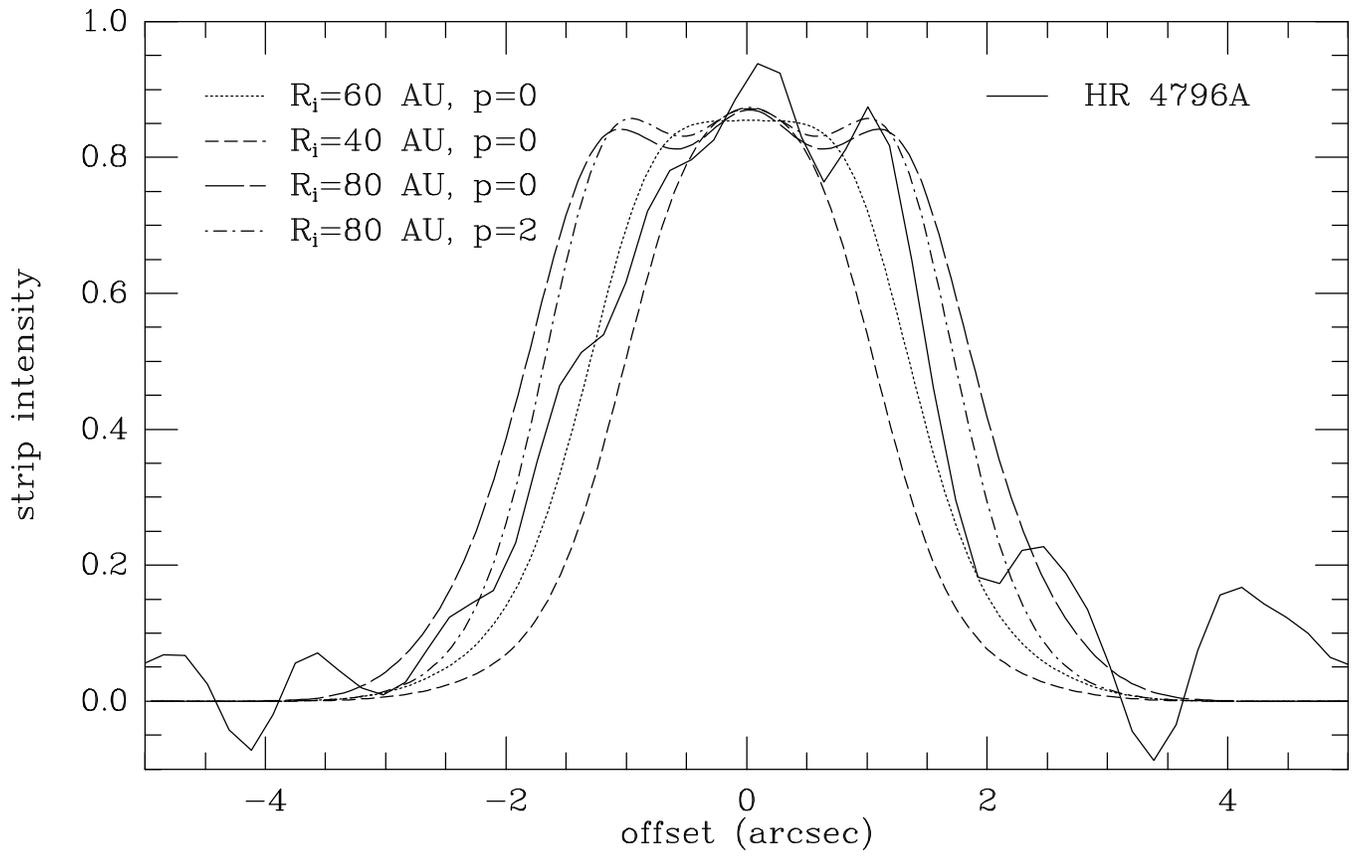
\end{document}